# Theoretical analysis of high-field transport in graphene on a substrate

Andrey Y. Serov[1,2], Zhun-Yong Ong[3], Massimo V. Fischetti[3], and Eric Pop[1,2,4]

*[1]Electrical and Computer Engineering, University of Illinois, Urbana-Champaign, IL 61801, USA*

*[2]Micro and Nanotechnology Lab, University of Illinois, Urbana-Champaign, IL 61801, USA*

*[3]Materials Science and Engineering, University of Texas at Dallas, TX 75080, USA*

*[4]Electrical Engineering, Stanford University, Stanford, CA 94305, USA*

We investigate transport in graphene supported on various dielectrics ($SiO_2$, BN, $Al_2O_3$, $HfO_2$) through a hydrodynamic model which includes self-heating and thermal coupling to the substrate, scattering with ionized impurities, graphene phonons and dynamically screened interfacial plasmon-phonon (IPP) modes. We uncover that while low-field transport is largely determined by impurity scattering, high-field transport is defined by scattering with dielectric-induced IPP modes, and a smaller contribution of graphene intrinsic phonons. We also find that lattice heating can lead to negative differential drift velocity (with respect to the electric field), which can be controlled by changing the underlying dielectric thermal properties or thickness. Graphene on BN exhibits the largest high-field drift velocity, while graphene on $HfO_2$ has the lowest one due to strong influence of IPP modes.

*[*]Contact: epop@stanford.edu

## I.    INTRODUCTION

Graphene is an interesting material for both fundamental and practical studies[1,2] due to its unusual linear band structure,[3] outstanding intrinsic mobility,[4] high thermal conductivity,[5] high transparency and mechanical strength.[6] Although graphene in suspended platforms demonstrates exceptionally high mobilities,[4] such test devices are limited to low carrier densities as intrinsic graphene is undoped and gating through an air or vacuum gap is ineffective. In order to achieve higher carrier densities for practical applications graphene must be placed on (or covered with) an insulating dielectric layer, where its transport characteristics are modified significantly, as such layers introduce additional scattering mechanisms associated with ionized impurities and dielectric remote phonons.[7]

There are several experimental and theoretical studies that have examined the role of ionized impurities on the formation of charge puddles and on charge transport in graphene at *low* electric fields, on insulating substrates.[8-11] Researchers have also found they can directly control impurity scattering in graphene by screening it with solvents of high dielectric constant.[12] However, the role of substrate phonons is difficult to quantify directly. Different theoretical approaches have been used to explain the role of substrate phonons on low- and high-field transport in graphene, with the role of screening by charge carriers in graphene left somewhat arbitrary: while some studies assumed no screening in the graphene layer,[13,14] others used static Thomas-Fermi screening models[15,16] and only recently a theory of dynamic screening in graphene, which takes into account the hybridization of the substrate phonons with graphene plasmons, has been introduced.[17,18] Although the measurement of the field-effect dependence of the *thermal* conductance with the substrate could elucidate the screening mechanism,[19] the appropriate data have not been published yet. This theory of substrate phonons (also known as interfacial plasmon-phonons, IPP) has so far only been applied to low-field transport in graphene and to a study of graphene-substrate thermal boundary conductance.[19]

High-field transport is an important regime, which is interesting for both fundamental physics and device design applications. Compared to the case of low electric fields, where the system is usually close to thermal equilibrium, physical effects at high electric fields are very different, as charge carriers driven out of equilibrium reach much higher energies,[20] which open more scattering channels and lead to significant power dissipation and accompanying thermal phenomena.[21-23] High electric fields can be achieved in high-performance or high-power analog transistors, which operate in the current saturation regime, typically at fields >1 V/μm. As the current saturation is an important metric which determines the transistor gain, a better understanding of velocity saturation and the role of the substrate is needed to advance the development of graphene-based electronics. As in the case of other materials and devices,[24,25] high current in graphene-based devices leads to lattice heating,[26] which must also be considered in realistic models.

In this work, we evaluate the role of the substrate during high-field transport in graphene. We consider the set of physical mechanisms introduced by the substrate such as remote phonon scattering, which is implemented with the theory of dynamic screening of charge carriers in graphene. We also include scattering with ionized scattering centers, such as fixed impurities or ionized interface traps, which can be introduced by the substrate and the graphene fabrication process, but are partially screened by substrate, too. We take into account the effects of self-heating and thermal coupling to the substrate as well. At large current densities in high-field transport, Joule self-heating from power dissipation can reach hundreds of kW/cm$^2$, increasing the temperature of the device and affecting its performance negatively, through a larger population of phonons which can scatter the carriers. In this paper, after outlying the theoretical foundation for our work, we benchmark the simulation results with experimental data for high-field drift velocity in graphene on SiO$_2$. Then we analyze and compare the roles of various physical mechanisms contributing to the high field transport, such as self-heating, impurity scattering and phonon scattering. Finally, we compare the high-field transport in graphene on several dielectrics such as SiO$_2$, HfO$_2$, Al$_2$O$_3$ and BN with and without self-heating.

## II. THEORY

### A. Transport model

In order to study high-field transport in graphene in contact with a substrate or dielectric, we need to take into account several physical effects such as carrier heating, lattice heating and various scattering mechanisms. In principle, this transport problem could be solved with the Monte-Carlo method, which is a powerful tool to treat various scattering mechanisms.[27,28] Instead, here we employ a hydrodynamic model that is computationally less demanding than the ensemble Monte Carlo method, especially when including self-consistently both self-heating effects and full inter-particle Coulomb interactions, which are important in graphene[29] at practical carrier densities >10$^{12}$ cm$^{-2}$. The carrier distribution function can be expressed as[30,31]

$$f_{\mathbf{k}}(\mathbf{v}_d, T_e) = \left[ \exp\left( \frac{E - \hbar\mathbf{v}_d \cdot \mathbf{k} - E_F}{k_B T_e} \right) + 1 \right]^{-1}, \qquad (1)$$

where $E$ is the carrier energy, $v_d$ is the average drift velocity, $\mathbf{k}$ is the carrier momentum (here with $x$- and $y$-components), $E_F$ is the Fermi level, $k_B$ is the Boltzmann constant and $T_e$ is the carrier (electron) temperature. We generically use electron subscripts, but the discussion could similarly apply to holes because the energy dispersion in graphene is symmetric. This distribution function [Fig. 1(a)] has several features. First, it is of the form expected from detailed-balance when inter-particle collisions are significant, as is the case in graphene at the carrier densities of interest. (A displaced Fermi-Dirac distribution minimizes the electron-electron scattering integral $S_{EE}$.) Second, the total current can be easily calculated as $|J| = env_d$, where $e$ is the elementary charge and $n$ is the carrier density. One of the drawbacks of this distribution function is that the carrier density $n$ is not only function of $E_F$, but also a function of $v_d$ and $T_e$.

The drift velocity $v_d$ and the electron temperature $T_e$ are determined from balancing momentum and energy gained by charge carriers from the electric field, with momentum and energy released through various scattering mechanisms.[30] We also introduce an insightful power dissipation and self-heating approach shown in Fig. 1(b). High-energy electrons in the graphene can scatter directly with graphene phonons or with substrate IPP modes. Heat generated in the former must cross the thermal boundary resistance (TBR) between graphene and the substrate. We recall that heat transfer from graphene to the substrate is characterized by a TBR[32] of the order $\sim 10^{-8}$ Km$^2$W$^{-1}$ with a weak dependence on the substrate material.[33] Power dissipated directly with substrate IPPs bypasses the TBR and the resulting heat is directly conducted into the substrate, towards the backside heat sink.

These heat flow pathways are captured by the thermal resistance network in Fig. 1(c), where the temperature rise across any two nodes is proportional to the product of the dissipated power and the thermal resistance between the nodes.[34] Most substrate dielectrics considered in this work have low thermal conductivity[35] (e.g. $\sim 1.4$ Wm$^{-1}$K$^{-1}$ for SiO$_2$ at room temperature) and they tend to dominate the heat flow path, although the precise balance of the thermal resistances also depends on the substrate thickness and interface quality or TBR. Using analytic models fitted to experimental data we also take into account the temperature dependence of the SiO$_2$ thermal conductivity[34] and the temperature dependence of the TBR between graphene and SiO$_2$.[36] We can neglect lateral heat flow here, as the experimental sample[34] which is used to benchmark our simulations is significantly larger ($4 \times 7$ μm) than the thermal healing length in graphene on 300 nm SiO$_2$ ($L_H \sim 0.2$ μm).[26,37-39] Lateral heat sinking to the contacts is negligible in devices of length $L \gg 3L_H$. The temperature profiles of shorter or narrower devices can be treated through finite element simulations[37] or sometimes through analytical solutions.[38,39]

Combining all these mechanisms we arrive at a multi-scale physics model, which can be described with a set of equations:

$$\Phi = \begin{cases} en\mathbf{F} + \sum_{\mathbf{k}} \mathbf{k} S(f_{\mathbf{k}}) \\ en\mathbf{F} \cdot \mathbf{v_d} + \sum_{\mathbf{k}} E_{\mathbf{k}} S(f_{\mathbf{k}}) \\ T_{gr} - T_{sub} + R_B \sum_{\mathbf{k}} E_{\mathbf{k}} \left[ S_{OP}(f_{\mathbf{k}}) + S_{AC}(f_{\mathbf{k}}) \right] \\ T_{sub} - T_0 + (R_{ox} + R_{Si}) \sum_{\mathbf{k}} E_{\mathbf{k}} S(f_{\mathbf{k}}) \end{cases} = 0 \qquad (2)$$

where $\mathbf{F}$ is the electric field, $S$ is the scattering integral, which includes graphene optical (OP) and acoustic (AC) phonons, parasitic (impurity) interface charge and substrate plasmon-phonon modes; $T_{gr}$ is the temperature of the graphene lattice, $T_{sub}$ is the temperature of a substrate at the interface with graphene, $T_0$ is the ambient temperature, $R_B$ is the graphene- substrate TBR, $S_{OP}$ and $S_{AC}$ are the scattering integrals due to graphene OP and AC phonons, respectively; $R_{sub}$ and $R_{Si}$ are the thermal resistance of the insulator and silicon substrate, respectively. The first and second equations in (2) describe the momentum and energy balance, while the third and fourth equations describe the thermal balance between graphene, underlying dielectric and heat sink.

We have a system of non-linear equations $\Phi(v_d, T_e, T_{gr}, T_{sub}) = 0$, which we solve using the Jacobian and the Newton-Raphson method, employing the simulation scheme shown in Fig. 2. First, we set the carrier density $n$ and electric field $F$, and the initial conditions for step 0: $v_d{}^0 = 0$, $T_e{}^0 = T_0$, $T_{gr}{}^0 = T_0$, $T_{sub}{}^0 = T_0$. Then at each step '$i$' we calculate the scattering integrals for momentum and power, and compute the calculation error $\Delta^i = \Phi(v_d{}^i, T_e{}^i, T_{gr}{}^i, T_{sub}{}^i)$ and Jacobian matrix $J^i$. Finally, we use the Newton-Raphson method to find $v_d{}^{i+1}$, $T_e{}^{i+1}$, $T_{gr}{}^{i+1}$, $T_{sub}{}^{i+1}$ by inverting the Jacobian and multiplying it by the vector of the error $\Delta^i$. Since the carrier density depends on $E_F$, $v_d$ and $T_e$, we also need to update $E_F$ at each step to keep the carrier density constant at the desired value.

### B. Carrier scattering mechanisms

We now turn to the explicit calculation of the scattering integrals. We use the linear band structure for graphene $E = \hbar v_F k$, where $v_F = 10^6$ m/s is the Fermi velocity, and we will justify this approximation later. The scattering rate with graphene OPs is described by the Fermi golden rule and deformation potential theory by summation over all possible phonon wavevectors, i.e.[13]

$$W_{k \to k'}^{OP} = \sum_q \frac{2\pi D_\Gamma^2}{\rho_c \omega_\Gamma} \Big[ \delta \big( \hbar v_F k - \hbar v_F k' + \hbar \omega_\Gamma \big) N_q(\omega_\Gamma) + \delta \big( \hbar v_F k - \hbar v_F k' - \hbar \omega_\Gamma \big) \big( 1 + N_q(\omega_\Gamma) \big) \Big]$$

$$+ \sum_q \frac{2\pi D_K^2 \big[ 1 - ss' \cos(\theta_{kk'}) \big]}{2 \rho_c \omega_K} \tag{3}$$

$$\times \Big[ \delta \big( \hbar v_F k - \hbar v_F k' + \hbar \omega_K \big) N_q(\omega_K) + \delta \big( \hbar v_F k - \hbar v_F k' - \hbar \omega_K \big) \big( 1 + N_q(\omega_K) \big) \Big]$$

where $q$ is the scattering wavevector and we used the deformation potential at $\Gamma$ and K points[40] $D_\Gamma =$ 7.9 eV/Å, $D_K = 13.9$ eV/Å, graphene density $\rho_c = 7.66 \times 10^{11}$ kg/cm$^2$, $k$ and $k'$ are the wavevectors of initial and final states, $N_q$ is the phonon population with the Bose-Einstein distribution, $\theta_{kk'}$ is the scattering angle, $s = \pm 1$ for electrons and holes, respectively. The phonon energies at the $\Gamma$ and K points are[41] $\hbar \omega_\Gamma = 196$ meV and $\hbar \omega_K = 161$ meV, respectively. The first and second term in Eq. (3) correspond to Brillouin zone center ($\Gamma$) and zone edge (K) phonons, respectively, where deformation potentials are computed using the GW method.[40]

We also take into account scattering with *inter*-valley transverse acoustic (TA) phonons using a simplified model[42] with the deformation potential $D_{TA} = 3.5$ eV/ Å and the phonon energy 124 meV, corresponding to K-point TA modes. We implement scattering with *intra*-valley acoustic phonons (AC) following the deformation potential formalism,[43] and using the deformation potential $D_{AC} = 25$ eV together with the Dirac overlap integral. This deformation potential for AC phonons has been fitted to the rigid ion model.[44] Although $D_{AC}$ is relatively high compared to the literature,[13,42,43] smaller deformation potentials lead to a much weaker velocity saturation especially at higher carrier density, which does not agree well with our experimental data.[34]

The scattering integrals can be calculated as

$$S(f_k) = -\sum_{k'} \Big[ W_{k,k'} f_k (1 - f_{k'}) - W_{k',k} f_{k'} (1 - f_k) \Big] \tag{4}$$

Scattering with ionized parasitic charge, including ionized impurities and charged traps at the graphene interfaces, is treated following the work by Adam *et al.* where the rate is[45]

$$P_{IMP} = \frac{2\pi}{\hbar} \left| \frac{1}{2\varepsilon_0} \frac{e^2}{\kappa (q + q_{TF})} \right|^2$$

$$\times \frac{1 + \cos(\theta_{kk'})}{2} \delta \big( \hbar v_F k - \hbar v_F k' \big) \tag{5}$$

where $\varepsilon_0$ is the vacuum dielectric constant, $\kappa = (1 + \kappa_{sub})/2$ is the effective dielectric constant[45] ($\kappa_{sub}$ is the dielectric constant of the substrate), and $q_{TF}$ is the static screening wavevector calculated in the Thomas-Fermi approximation.

A substrate scattering mechanism which is usually neglected or treated trivially is the scattering with IPP modes formed from the hybridization of substrate OPs with graphene plasmons.[17] Since the theory is quite lengthy, we discuss it here very briefly and refer the interested reader to Refs. [17,46] for a more complete description. The plasmons in graphene couple electromagnetically to the two OPs in the dielectric, forming three interfacial plasmon-phonon branches (IPP1, IPP2 and IPP3), as shown in Fig. 3. In the long wavelength limit, one IPP branch converges to the free plasmon dispersion in graphene, and the two other branches converge with the bare substrate ($SiO_2$ in Fig. 3) transverse optical (TO) phonon branches. There are two discontinuities in the dispersion for IPP1 and IPP2: the first jump occurs as a result of the discontinuity in the density-density response function when the phonon branch crosses the line $E = \hbar v_F k$; the second jump occurs at a higher wavevector as a result of Landau damping when the IPP3 branch crosses the line $E = \hbar v_F k$. We use 3 Å as the spacing between graphene and the dielectric. Our model[17,46] differs from the more commonly used static screening model[15,16] in that the screening of the remote phonon at long wavelengths does not diverge to infinity. Hence, in our model the IPP coupling with electrons diverges as $q \to 0$, while in the more commonly used static screening model the coupling strength remains finite as $q \to 0$. At low carrier density, the scattering by long wavelength modes is dominant and thus, carrier mobility is more strongly degraded by electron-IPP scattering.

## III. RESULTS AND DISCUSSION

### A. Graphene on $SiO_2$

We first evaluate our model by comparing and benchmarking its simulation results to extracted experimental data[34] for exfoliated graphene on $SiO_2$. We observe that the low-field mobility has a weak temperature dependence below 350 K[10,34] which indicates that scattering with interface charge centers is the dominant momentum relaxation mechanism at low fields. The low-field mobility also decreases with higher gate bias, which is a signature of elastic scattering or different interface trap occupancy for different gate biases $V_G$.[47] We find that AC phonon scattering is insufficient to describe the decrease of low-field mobility with higher $V_G$, and we use the density of ionized impurities (or traps) as a fitting parameter, which depends on $V_G$. As $V_G$ varies, different numbers of interface traps are occupied, which is related to hysteresis experimentally observed for graphene on various dielectrics.[48,49] Change in the occupation of interface traps leads to both a change in carrier density and a change in the density of ionized scatterers at the interface. Although short-range defect scattering with grain boundaries and line defects can also lead to a decrease of mobility at high $V_G$, these are common in graphene grown by chemical vapor deposition, not in the (more pristine) exfoliated samples we use to benchmark our simulations here. A similar discussion of phenomena related to the variation of trapped impurity charge is found in Refs. 44 and 47.

The plot of the simulation and experimental data for drift velocity vs. electric field is shown in Fig. 4(a) and (b) for the base temperatures $T_0 = 300$ K and 80 K, respectively. We use the interface parasitic charge density as a fitting parameter for different gate biases. We estimate the carrier concentration to be $n = C_{ox}(V_G - V_0)/e - [n_{imp}(V_G) - n_{imp}(V_0)]$, where $C_{ox}$ is the oxide capacitance, $V_0$ is the Dirac voltage and $n_{imp}$ is the density of ionized scattering centers at the interface as a function of $V_G$, which consists of fixed and trapped charges. The ionized charge densities extracted for $T_0 = 300$ K are shown in Table 1. As we can observe in Fig. 4, the drift velocity saturates at electric fields higher than 1 V/μm and exhibits a negative slope at even higher fields, which we attribute to lattice heating. We analyze the contribution of various scattering mechanisms in Fig. 5(a)-(c), where three different carrier densities ($n = 1.3 \times 10^{12}$, $2.5 \times 10^{12}$ and $3.8 \times 10^{12}$ cm$^{-2}$) are considered. We see that impurity scattering dominates momentum relaxation at lower electric fields, while IPP modes are more significant at higher fields. The contribution of graphene OP and AC phonons increases with carrier density while the relative contribution of IPP modes decreases with carrier density, which is caused by screening. Figure 5(d) shows the dependence of $T_e$ and $T_{sub}$ on the electric field at $n = 1.3 \times 10^{12}$ cm$^{-2}$ and $n = 2.5 \times 10^{12}$ cm$^{-2}$. As the electric field strength increases, the higher power dissipated in the structure leads to greater device heating and higher temperatures. At higher carrier density, the corresponding power density is also larger and the lattice heats up more. Since the probability of phonon scattering increases with lattice temperature, the temperature rise from self-heating automatically leads to greater momentum and energy loss to the lattice. In other words, like silicon-on-insulator (SOI) structures,[50] the saturation behavior is strongly dependent on self-heating which is modulated by the thermal resistance of the substrate.

## B. Role of self-heating

Figure 6(a) shows the saturation behavior for structures with different SiO$_2$ thicknesses, which controls lattice heating since the dielectric constitutes the largest part of the thermal resistance of the system. (We note this is the case for large devices with lateral dimensions $L$ and $W \gg t_{sub}$ and uniform electric field. However, in smaller, approximately sub-0.3 μm graphene devices the lateral heat spreading to the contacts also becomes important and must be included in a complete thermal analysis.[37,38]) A thinner $SiO_2$ corresponds to lower thermal resistance, better thermal grounding, less lattice heating and hence a higher $v_d$. The velocity saturation is much weaker when the effective $SiO_2$ thickness is less than 100 nm, and the high-field behavior for graphene supported by 30 nm of $SiO_2$ (on Si) is very close to the perfect heat sinking case. The saturation behavior analyzed at two different carrier densities [Fig. 6(a) and (b)] also shows that these two cases look very similar qualitatively. Since thermal conductivities of amorphous $SiO_2$, $HfO_2$ and $Al_2O_3$ are similar[51] (at least at room temperature), 30 nm of $SiO_2$ corresponds to about the same thickness for $HfO_2$ or $Al_2O_3$ in terms of thermal resistance. However, hBN has a higher and anisotropic thermal conductivity[52,53] which can lead to better cooling even for somewhat thicker dielectric layers.

### C. Role of band nonlinearity

Until now, we have assumed the simple linear band structure of graphene. Given that the application of a strong electric field leads to a skewed non-equilibrium carrier distribution [see Fig. 1(a)], we now check how deviations from the linear band assumption are likely to affect our results. Figure 7(a) shows the calculated graphene dispersion along the K-M axis, where non-linearity is the strongest, using the tight-binding model.[54] The energy dispersion begins to deviate slightly from the linear approximation for wavevectors $k > 2$ nm$^{-1}$, but is still close to linear for $k < 4$ nm$^{-1}$. These correspond to the carrier energy range from 1.3 to 2.6 eV. Following our calculation of the distribution function, as shown in Fig. 7(b) and Fig. 7(c), we choose the set of parameters $v_d = 0.5 v_F$, $T_e = 650$ K and $E_F = 0.2$ eV (corresponding to $n = 5.4 \times 10^{12}$ cm$^{-2}$), which on the one hand describe a highly non-equilibrium distribution, but on the other hand represent an upper bound in a given simulation with the linear band structure. Using these parameters we plot the distribution function shown in Fig. 7(b) and along the $k_x$ axis in Fig. 7(c), where we can see that $f(k) < 10^{-5}$ for $k > 2$ nm$^{-1}$ and $f(k) < 10^{-10}$ for $k > 4$ nm$^{-1}$, which means that the number of carriers with energy high enough to reach the nonlinear part of the energy band is negligible. The number of charge carriers reaching the nonlinear region is very small because of several factors: first, electron-electron scattering reshapes the distribution function limiting the carrier energy and second, scattering with the substrate reduces the lifetime of the high energy carriers, leading to significant momentum and energy loss to the substrate.

### D. Role of impurity scattering

We now examine the effect of the interface impurity scattering on velocity saturation. We calculate the drift velocity as a function of electric field for two different carrier densities ($n = 10^{12}$ cm$^{-2}$ and $n = 3 \times 10^{12}$ cm$^{-2}$) and two different impurity concentrations ($n_{imp} = 10^{11}$ cm$^{-2}$ and $n_{imp} = 6 \times 10^{11}$ cm$^{-2}$) as shown in Fig. 8(a). We can see that while the low field mobility is determined primarily by the impurity concentration and depends weakly on the carrier density in the given range, the high-field behavior depends more strongly on the carrier density than on the impurity concentration. This highlights the important role of phonon scattering at higher fields: as the distribution function becomes more skewed and heats up, it is easier to emit high energy phonons, as seen in Fig. 5(a) to (c). In Fig. 8(b), we show the dependence of high-field velocity $v_{HF}$, defined as the drift velocity at electric field $F = 1$ V/µm, on the carrier density for three different impurity concentrations. We observe that at lower carrier densities $v_{HF}$ depends more strongly on impurity concentration but at higher carrier density $v_{HF}$ demonstrates an asymptotic behavior almost independent on impurity charge, which can be explained by the stronger screening of ionized impurities at higher carrier densities and the greater role of phonon scattering.

### E. Role of screening interface modes

As discussed above, we consider scattering with hybrid IPP modes, which incorporate the effect of dynamic screening[17] through coupling between the surface phonons and the graphene plasmons, and scattering with unscreened bare surface phonon modes.[13] Screening of surface modes in general leads to weaker coupling of the substrate phonons to the graphene charge carriers, resulting in higher mobility and lower thermal boundary conductance between the graphene and substrate.[19] The importance of such considerations for high-field behavior are shown in Fig. 9(a), where we note that unscreened surface polar phonon modes (SPP) lead to a lower mobility and notably lower velocity at high fields. We also plot $v_{HF}$ extracted at $F = 1$ V/µm as a function of carrier density in Fig. 9(b), where, despite qualitatively similar dependence on carrier density, $v_{HF}$ calculated with SPP modes is noticeably lower than $v_{HF}$ calculated with IPP modes.

### F. Comparison of different dielectrics

Finally, we compare the high-field transport behavior in graphene on various dielectrics. We use different impurity densities to reproduce experimentally extracted low-field mobilities[55,56] of graphene on BN (~14000 cm$^2$V$^{-1}$s$^{-1}$), on SiO$_2$ (8200 cm$^2$V$^{-1}$s$^{-1}$), on Al$_2$O$_3$ (7500 cm$^2$V$^{-1}$s$^{-1}$) and on HfO$_2$ (3600 cm$^2$V$^{-1}$s$^{-1}$), all at room temperature. We compare the high-field transport at carrier density $n = 2 \times 10^{12}$ cm$^{-2}$ and thermal resistance equivalent to 300 nm of SiO$_2$ in Fig. 10(a), followed by the same case with zero thermal resistance in Fig. 10(b). The dielectric properties of the substrate phonon modes are used as described in the literature.[13,19] We find that in the case with strong lattice heating ($t_{sub}$ = 300 nm) in Fig. 10(a), transport in graphene on all dielectrics exhibits negative differential drift velocity. Although the lower field mobility value can vary with the dielectric, at higher fields ($F >$ 1.5 V/µm) the drift velocities are closer to each other. In the case of effective heat sinking ($t_{sub}$ = 0 nm), the drift velocity does not saturate for all dielectrics in the given field range.

Graphene on $HfO_2$ has both the lowest mobility and lowest high-field drift velocity due to higher impurity density and low-energy substrate phonons in $HfO_2$, which couple strongly to the charge carriers in graphene. Graphene on both $SiO_2$ and $Al_2O_3$ have similar mobilities[55] and high-field drift velocities. Graphene on BN has a superior mobility[56] and high-field drift velocity due to its cleaner interface and high-energy dielectric phonons with weak coupling to charge carriers because of its relatively lower dielectric constant. This confirms BN as an optimal substrate material for low and high-field applications of graphene.

We also plot the drift velocity $v_d$ as a function of electric field with zero impurity density in Fig. 10(c,d) with and without self-heating. These correspond to the intrinsic, ideal behavior of clean graphene on clean dielectric. In this case the behavior is qualitatively similar with higher drift velocities overall but smaller difference between BN and $SiO_2$ as both have similar dielectric properties.

## IV. CONCLUSION

This study represents the first time that the high-field behavior in graphene on a substrate was investigated taking into account intrinsic graphene properties, hybrid interface plasmon-phonon modes and lattice self-heating. Wherever available, we benchmarked our theoretical velocity-field dependencies against available experimental data in a range of temperatures, electric fields and carrier densities. We examined high-field transport in graphene on various dielectrics such as $SiO_2$, $Al_2O_3$, BN and $HfO_2$. We found that while the low-field transport is determined by ionized impurities at the interface, the high-field behavior is determined by scattering with interfacial plasmon-phonon modes and a small contribution of graphene intrinsic phonons. Since phonon scattering is dominant at higher fields, lattice self-heating strongly affects the high-field behavior and can lead to a negative slope in drift velocity vs. electric field especially for devices on thicker and thermally resistive substrates. If, on the other hand, self-heating is completely suppressed, the drift velocity does not necessarily saturate at high field on any of the dielectrics investigated. (Cognizant of this, we often used the notation $v_{HF}$ instead of $v_{sat}$.) We have also shown that the dynamic screening of substrate phonon modes is important as it leads to higher mobility and higher drift velocity especially at higher carrier densities. Finally, we found that graphene on BN has the highest drift velocity at high fields and graphene on $HfO_2$ the lowest one, while graphene on $SiO_2$ and on $Al_2O_3$ are in between these ranges.

## ACKNOWLEDGEMENT

The authors acknowledge valuable discussions with V.E. Dorgan, J.-P. Leburton, U. Ravaioli and E. Rosenbaum. This work was supported in part by the National Science Foundation (NSF) CAREER (ECCS-0954423) award (E.P.), by the Air Force FA9550-10-1-0082 award, by SONIC, one of six STARnet Centers sponsored by MARCO and DARPA, and by South West Academy of Nanoelectronics (SWAN) and the Nanoelectronic Research Initiative (NRI) Task 4.3 Theme 2400.011.

| $V_G - V_0$ (V) | Carrier density $n$ (cm$^{-2}$) | Ionized charge at the interface $n_{imp}$ (cm$^{-2}$) |
| --- | --- | --- |
| 0 | | $2.63 \times 10^{11}$ |
| 23.5 | $1.3 \times 10^{12}$ | $6.1 \times 10^{11}$ |
| 43.5 | $2.5 \times 10^{12}$ | $8.5 \times 10^{11}$ |
| 63.5 | $3.8 \times 10^{12}$ | $10.5 \times 10^{11}$ |

Table 1: Impurity densities used to fit experimental data shown in Fig. 4(a).

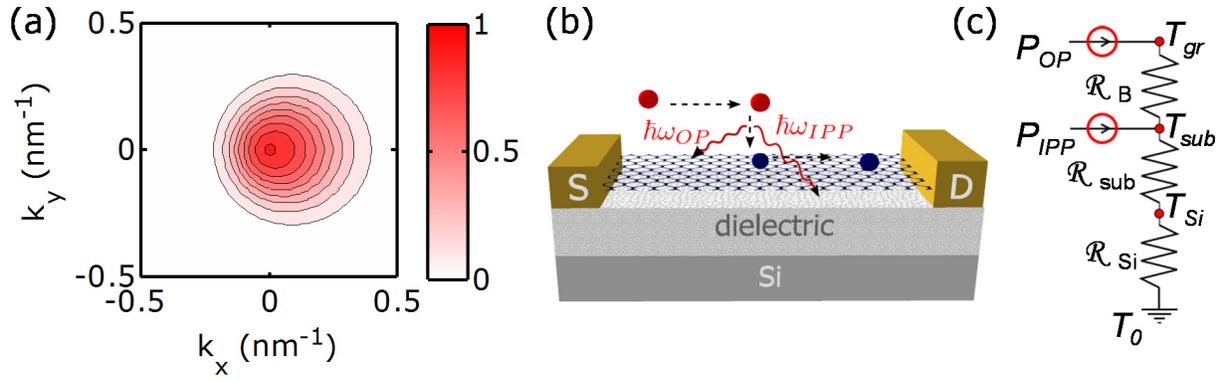

FIG 1. (a) Distribution function given by eq. (1) at $v_d = 0.3v_F$, $T_e = 450$ K, $E_F = 0.1$ eV. (b) Schematic of graphene on a substrate illustrating heat dissipation. Optical phonons (OP) dissipate power into the (acoustic) graphene lattice, which then must couple to the substrate through the thermal boundary resistance. Interfacial plasmon-phonon modes (IPP) dissipate power by a "shortcut" directly into the dielectric substrate. (c) Thermal network demonstrating heat propagation in the structure from initial heat dissipation through phonons to the heat sink.

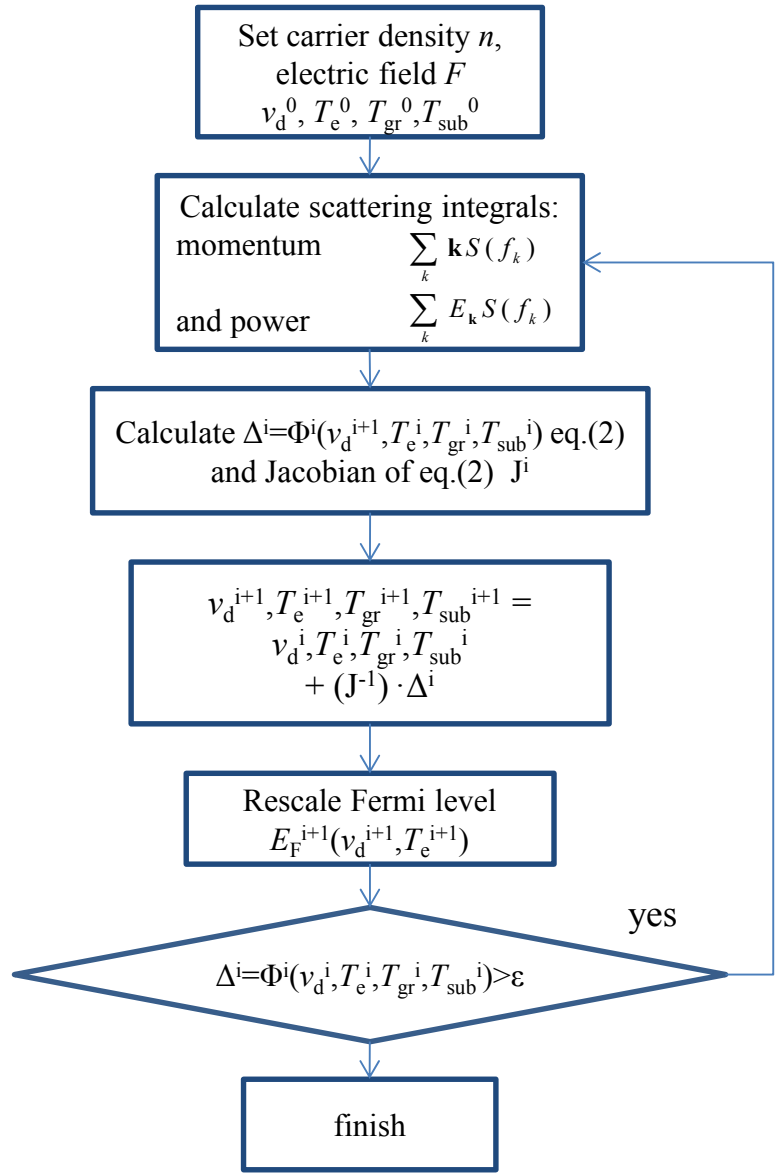

FIG 2. Calculation scheme. We use the Newton-Raphson method to solve a system of four non-linear equations (eq. 2) with four unknown ($v_d$, $T_e$, $T_{gr}$, $T_{sub}$).

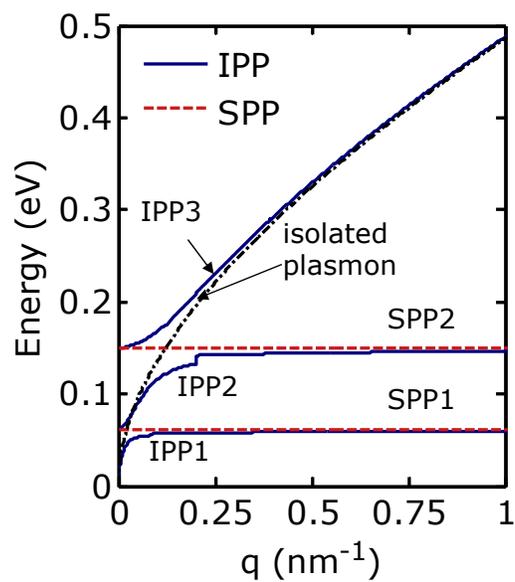

FIG 3. Interface plasmon-phonon modes (IPP1, IPP2, IPP3) for graphene on SiO$_2$ with carrier density $n$ = $2 \times 10^{12}$ cm$^{-2}$. Isolated plasmon mode in graphene is also shown. SPP1, SPP2 are surface modes calculated without screening, which correspond to dielectric transverse optical phonon modes TO1 and TO2 in SiO$_2$.

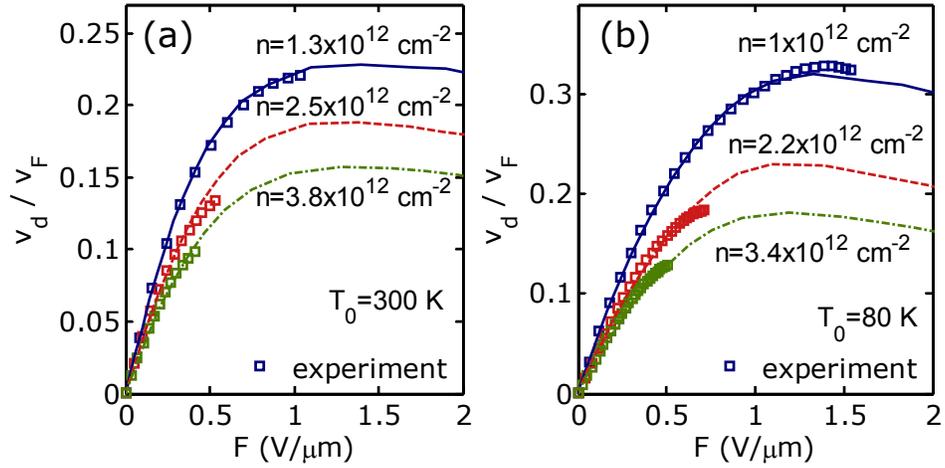

FIG 4. Drift velocity $v_d$ normalized by Fermi velocity $v_F$ as a function of electric field: simulations (lines) and experimental data[34] (symbols) for various carrier densities $n$ on SiO₂. (a) Ambient temperature $T_0 = 300$ K and (b) $T_0 = 80$ K, but note that the sample self-heats at high field, which is also responsible for the negative differential drift velocity (compare with Fig. 6 and Fig. 10).

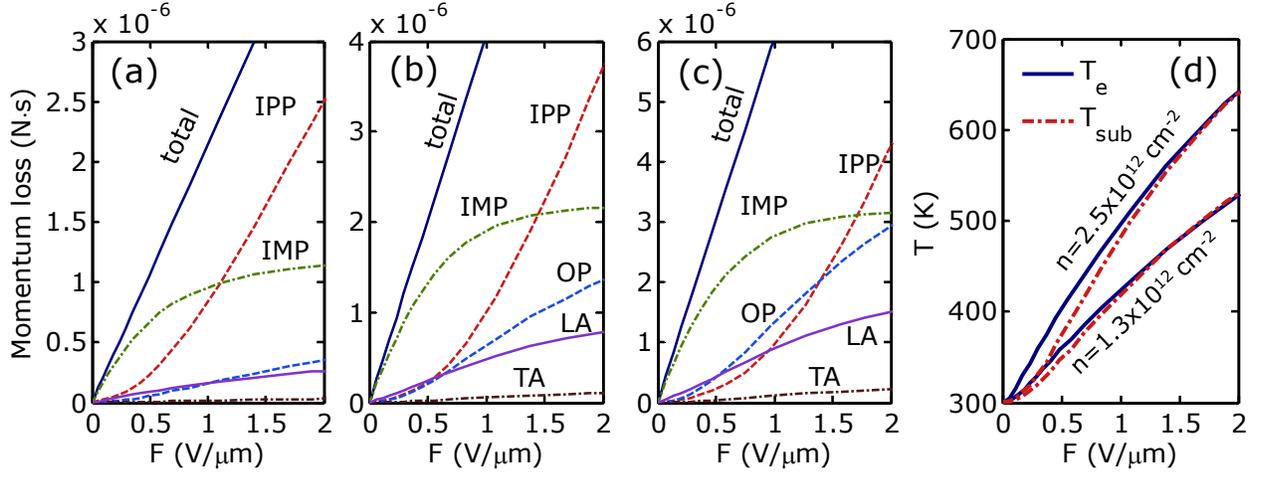

FIG 5. Momentum loss due to various scattering mechanisms in graphene on SiO₂ for three carrier densities: (a) $n = 1.3 \times 10^{12}\,\text{cm}^{-2}$, (b) $n = 2.5 \times 10^{12}\,\text{cm}^{-2}$, (c) $n = 3.8 \times 10^{12}\,\text{cm}^{-2}$. Please see text for discussion of IPP, IMP, OP, LA and TA scattering mechanisms. (d) Electron temperature $T_\text{e}$ and substrate temperature $T_\text{sub}$ as a function of electric field for two carrier densities, as labeled.

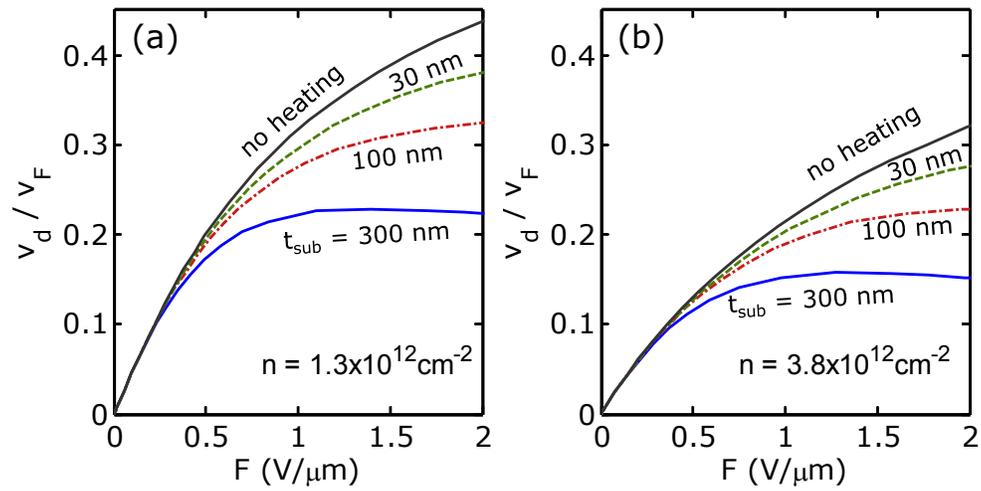

FIG 6. The role of lattice self-heating on high-field drift velocity in graphene on different oxide (SiO$_2$) thicknesses $t_{sub}$ and different carrier densities at $T_0 = 300$ K. Thicker oxides correspond to higher thermal resistance and greater lattice heating. (a) Carrier density $n = 1.3 \times 10^{12}$ cm$^{-2}$, (b) $n = 3.8 \times 10^{12}$ cm$^{-2}$.

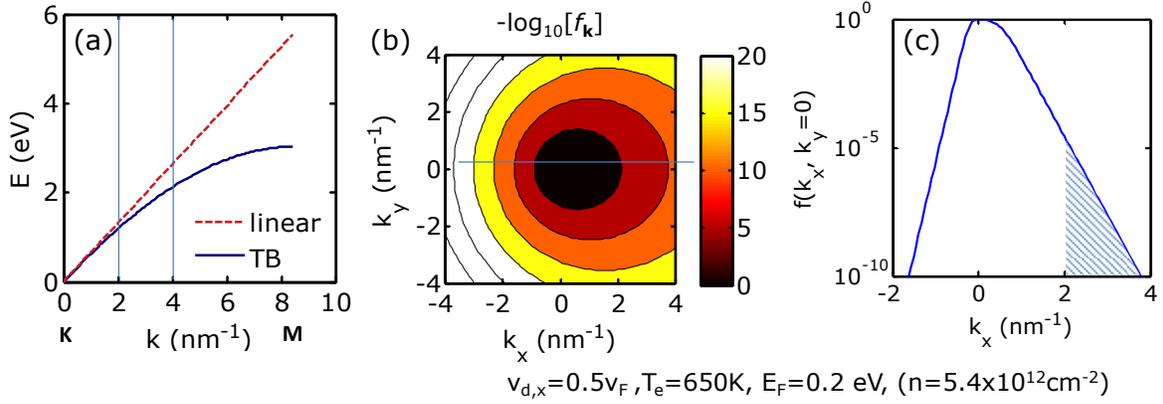

FIG 7. (a) Comparison of linear energy dispersion of graphene (dashed) with that calculated from the tight-binding approximation (TB), both shown along K-M direction with the strongest non-linearity. The dispersion begins to diverge from the linear approximation for $k > 2$ nm$^{-1}$ but remains close to linear for $k < 4$ nm$^{-1}$, which corresponds to the carrier energy range from 1.3 to 2.6 eV. (b) Distribution function [as $\log_{10}(f)$] calculated for $v_{d,x} = 0.5v_F$, $T_e = 600$ K, $E_F = 0.2$ eV ($n = 5.4 \times 10^{12}$ cm$^{-2}$). (c) Distribution function $f$ along the $k_x$ direction ($k_y = 0$) showing that $f < 10^{-5}$ for $k_x > 2$ nm$^{-1}$ and $f < 10^{-10}$ for $k_x > 4$ nm$^{-1}$. Thus, the linear dispersion approximation of graphene appears sufficient for all fields and carrier densities simulated in this work.

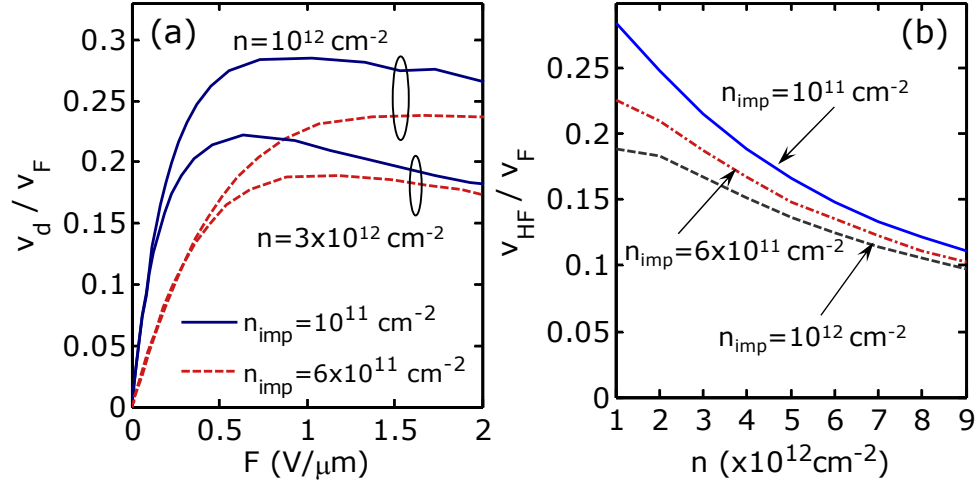

FIG 8. The role of impurity scattering on high-field drift velocity of graphene. (a) Drift velocity $v_d$ normalized by $v_F$ vs. electric field for two carrier densities, $n = 10^{12}$ cm$^{-2}$ and $n = 3 \times 10^{12}$ cm$^{-2}$ and two impurity densities $n_{imp} = 10^{11}$ cm$^{-2}$ and $n_{imp} = 6 \times 10^{11}$ cm$^{-2}$. (b) Drift velocity at high electric field ($v_{HF}$ at $F = 1$ V/µm) as a function of carrier density for three impurity densities $n_{imp}$.

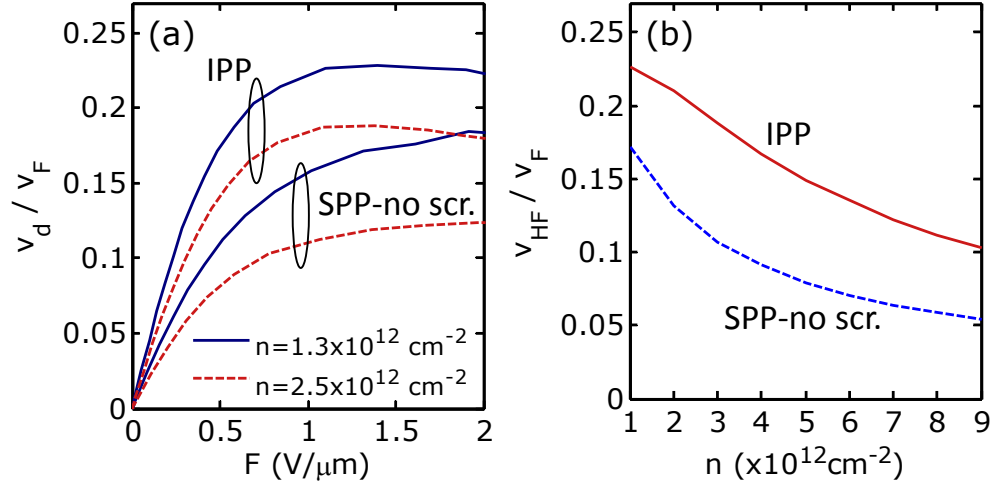

FIG 9. Role of IPP modes vs. surface modes without screening for high-field drift velocity in graphene on SiO$_2$, at room temperature. (a) Drift velocity $v_d$ normalized by $v_F$ vs. electric field for carrier densities $n = 1.3 \times 10^{12}$ cm$^{-2}$ and $n = 2.5 \times 10^{12}$ cm$^{-2}$, corresponding to Fig. 3. (b) Drift velocity computed at high electric field ($v_{HF}$ at $F = 1$ V/$\mu$m) as a function of carrier density including IPP modes and SPP modes without screening. (Impurity density $n_{imp} = 6 \times 10^{11}$ cm$^{-2}$).

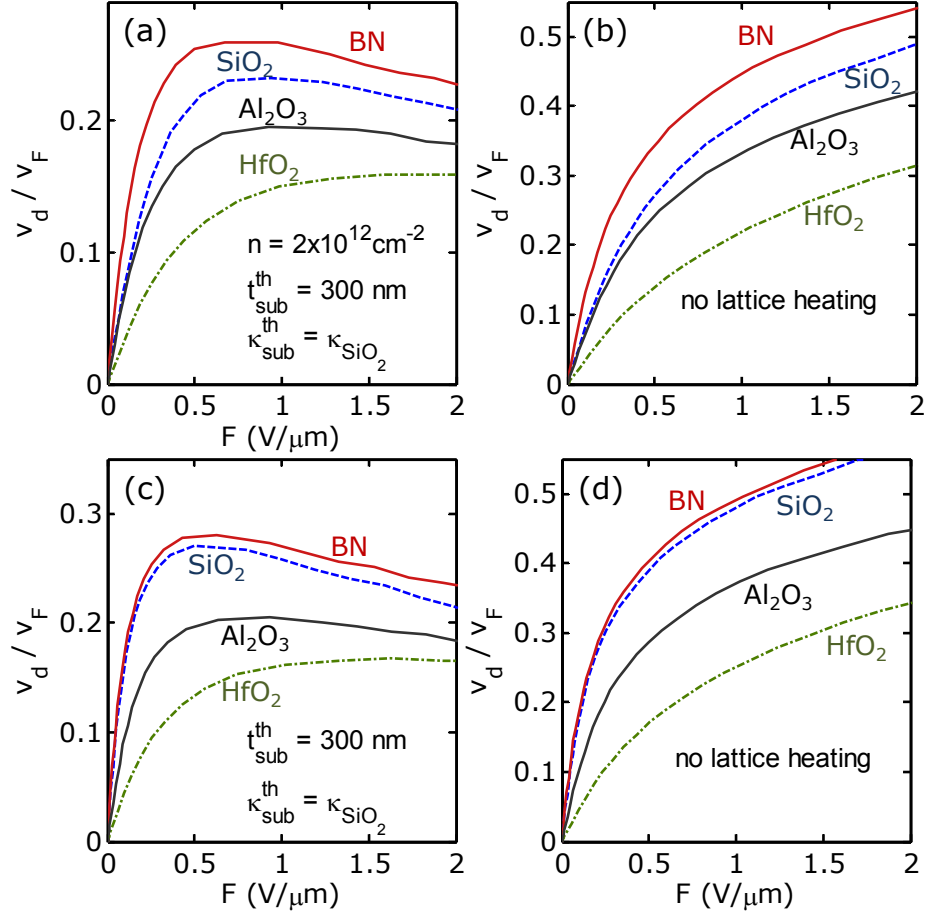

FIG 10. Role of impurities and IPP modes on high-field drift velocity in graphene on various dielectrics: BN, SiO₂, Al₂O₃, HfO₂. Impurity density in (a) and (b) is set to match the range experimental mobilities[55,56] (BN: 14000 cm²V⁻¹s⁻¹, SiO₂: 8200 cm²V⁻¹s⁻¹, Al₂O₃: 7500 cm²V⁻¹s⁻¹, HfO₂: 3600 cm²V⁻¹s⁻¹ ). (a) The case with strong self-heating, corresponding to a substrate with thermal resistance equivalent to that of 300 nm of SiO₂ for all substrates. Negative differential velocity is observed due to self-heating at fields > 1 V/μm for all substrates except HfO₂. (b) The ideal case without self-heating showing only weak saturation of the drift velocity is expected. (c) Oxide thickness is 300 nm corresponding to strong self-heating, but assuming ultra-clean sample, $n_{imp} = 0$. (d) Calculated results for best "intrinsic" high-field properties with no self-heating and no impurities ($t_{sub} = 0$, $n_{imp} = 0$). In this case we expect the highest drift velocity, but only weak saturation.